\documentclass[journal]{IEEEtranTCOM}

\normalsize

\usepackage{graphicx}

\usepackage{amsmath}
\usepackage{algorithm}
\usepackage[noend]{algpseudocode}
\usepackage{array}
\usepackage{multirow}
\usepackage{fancyhdr}

\begin{document}
\title{Optical OFDM Waveform Construction by Combining Real and Imaginary Parts of IDFT}

\author{Gokce~Hacioglu,~\IEEEmembership{}
        Kadir~Turk,~\IEEEmembership{}
        and~Cenk~Albayrak~\IEEEmembership{}% <-this % stops a space
\thanks{The work of K. Turk and C. Albayrak was supported by the Scientific and Technological Research Council of Turkey (TUBITAK) Project Number 215E308.}% <-this % stops a space
\thanks{G. Hacioglu and K. Turk are with the Department
of Electrical and Electronics Engineering, Karadeniz Technical University, 61080 Trabzon,
Turkey (e-mail: \{gokcehacioglu,kadir\}@ktu.edu.tr).}% <-this % stops a space
\thanks{C. Albayrak is with the Department
of Electronics and Communication Engineering, Karadeniz Technical University, 61830 Trabzon,
Turkey (e-mail: albayrak.cenk@ktu.edu.tr).}}% <-this % stops a space

\markboth{T\MakeLowercase{his work has been submitted to the} IEEE \MakeLowercase{for possible publication.} C\MakeLowercase{opyright may be transferred without notice, after which this version may no longer be accessible.}}{}

\maketitle

\begin{abstract}
%\boldmath
In optical communication systems, orthogonal frequency division multiplexing (OFDM) is widely used to combat inter-symbol interference (ISI) caused by multipath propagation. Optical systems which use intensity modulation and direct detection (IM/DD) can only transmit real valued symbols, but the inverse discrete Fourier transform (IDFT) or its computationally efficient form inverse-fast Fourier transform (IFFT) required for the OFDM waveform construction produces complex values. Hermitian symmetry is often used to obtain real valued symbols. 
For this purpose, some trigonometric transformations such as discrete cosine transform (DCT) are also used, however these transformations can eliminate the ISI only under certain conditions.
In this paper, we propose a completely different method for the construction of OFDM waveform with IFFT to obtain real valued symbols by combining the real and imaginary parts (CRIP) of IFFT output electrically (E-CRIP) or optically (O-CRIP). 
Analytical analysis and simulation works are presented to show that compared to the Hermitian symmetric system, the proposed method slightly increases the spectral efficiency, eliminates ISI, significantly reduces the amount of needed calculation and does not effect the error performance.
In addition, the O-CRIP method is less affected by clipping noise that may occur due to the imperfections of the transmitter front-ends. 
\end{abstract}


\IEEEpeerreviewmaketitle

\section{Introduction}\label{introduction}

\IEEEPARstart{O}{rthogonal} frequency division multiplexing (OFDM) is a powerful technique commonly used to combat inter-symbol interference (ISI) caused by multipath propagation in high data rate communication systems \cite{Goldsmith}. 
To remove ISI, OFDM converts the broadband signal into multiple narrow-band signals using the inverse discrete Fourier transform (IDFT) or its computationally efficient form inverse-fast Fourier transform (IFFT) and uses the cyclic prefix (CP). In this way, it creates parallel orthogonal channels narrower than coherence bandwidth.

OFDM is also widely used in optical communication systems. The IFFT process generates complex valued samples, but optical communication systems which use intensity modulation and direct detection (IM/DD) can only transmit real valued and non-negative symbols \cite{Goldsmith}. 
This prevents the direct use of OFDM for IM/DD optical communication systems. 
Hermitian symmetry method is frequently used in the literature to obtain real valued samples \cite{Armstrong1, Armstrong2}. In this method, complex valued symbols arranged according to the Hermitian symmetry structure are transformed into real valued symbols at the IFFT output. One approach to obtain non-negative symbols is to add a direct current (DC) bias to obtained real valued symbols. 

On the other hand, some trigonometric transforms which produce real valued samples may be used for OFDM waveform construction \cite{Mandyam}. Discrete cosine transform (DCT) is one of the transforms which produce real valued samples and there are numerous researches on using DCT for OFDM \cite{Ouyang1, Dhahir}. Fast OFDM is another method that uses trigonometric transformations to generate real valued samples \cite{Zhao}.
Fast OFDM reduces the subcarrier spacing twice while maintaining orthogonality between subcarriers in case of real valued modulation \cite{Xiong, Long}. 
However, OFDM with DCT, Fast OFDM and similar methods can eliminate the ISI only under certain conditions~\cite{Ouyang2}. 

In this paper, we propose a completely different method to produce real valued OFDM waveform with IFFT. 
In this method, instead of complex valued symbols as in the Hermitian symmetry method, the IFFT of real mapped symbols is calculated. Then the real and imaginary parts of the complex valued symbols generated at the IFFT output are simply summed to obtain real valued symbols. 
At the receiver side, after removing CP, real and imaginary parts of FFT outputs are subtracted from each other to recover original data. This method will be referred to as combination of real and imaginary parts (CRIP). Analyzes show that, the proposed CRIP method can completely remove ISI by using a single tap equalizer just like Hermitian symmetry method does. 

In order to produce a real valued symbol sequence with an average value of zero at the IFFT output, Hermitian symmetry method has to set the first and middle subcarriers of OFDM frame to zero and can not employ these two subcarriers for data transmission. 
Therefore, an OFDM frame with $N$ subcarriers constructed according to Hermitian symmetry structure can only transmit $(N-2)/2$ complex valued symbols which are equivalent to $N-2$ real valued symbols. 
However, the proposed CRIP method can employ middle subcarrier for data transmission and only needs to set the first symbol to zero to obtain the real valued sequence with zero average value. Therefore, CRIP method can transmit $N-1$ real valued symbols with one OFDM frame which means $1$ real valued symbol more than Hermitian symmetry.
But, note that even if the first subcarrier is not set to zero in the CRIP method, real valued symbols can still be obtained since real and imaginary parts of complex valued symbols of the IFFT output are simply summed. 
In this case, the mean value of the obtained symbol sequence will lightly differ from zero by $s_0/N$ where $s_0$ represents first data symbol of the sequence and it is real valued.
Numerical analysis shows that using the first subcarrier to carry data causes almost no error performance degradation. Then, an OFDM frame can transmit $N$ real valued symbols, which means slightly higher spectral efficiency by $N/(N-2)$ times than Hermitian symmetry method.
In addition to this advantage, to obtain real valued OFDM frame according to the proposed CRIP method significantly decreases the required computational complexity compared to the Hermitian symmetry method and even trigonometric transform based methods.

In recent years, the use of light-emitting diodes (LEDs), which have longer-lasting, higher energy efficiency and environmentally friendly lighting compared to traditional lighting devices, has become quite common. The widespread use of LEDs has attracted great interest in visible light communication (VLC), a type of short-range optical wireless communication that uses LEDs as transmitters \cite{VLC-6G}. 
In case of the optical OFDM waveform to be used in VLC systems, real valued signal obtained at the IFFT output is amplified, and the necessary DC bias is added in order to have non-negative symbols. The amplifier gain and DC bias level are adjusted to use the active regions of the LEDs. 
But, imperfections of the transmitter front-ends due to the use of off-the-shelf components and operating conditions of transmitter may shift the adjusted amplifier gain and DC bias level and result signal clipping \cite{dimitrov2012clipping}.
In a VLC system that use the proposed CRIP method, the real and imaginary parts of the IFFT output can feed individual LED chips without being combined before transmission. 
In this case, real and imaginary parts that are transformed into optical signals are combined in the transmission medium due to the nature of the light.  
Optical combination of real and imaginary parts will be referred to as optical CRIP (O-CRIP) in the article, and the combination at IFFT output as electrical CRIP (E-CRIP). 
Note that due to the low power of a LED chip, multiple LED chips are used together in LED lighting and VLC systems.  
We show by analytical and simulation works that O-CRIP method decreases the clipping noise caused by the imperfect front-ends, and the related system has significantly better error performance than the Hermitian symmetry method has. 
%The improvement effect of the clipping noise reduction on the system performance is demonstrated. \\

\indent The rest of the paper is organized as follows. 
Section II presents the system model we consider for theoretical analysis and shows that the proposed CRIP methods completely eliminate ISI as Hermitian symmetry does. 
Section III analyzes the clipping noise that occurs in the presence of clipping in the communication system for the E-CRIP, O-CRIP and Hermitian symmetry methods. 
Section IV gives the computational complexity of the both CRIP methods, Hermitian symmetry and trigonometric transformation based methods. 
Section V compares the methods in an example of typical VLC system.
Finally, Section VI concludes the paper.

%%%%%%%%%%%%%%%%%%%%%%%%%%%%%%%%%%%%%%%%%%%%%%%%%%%%%%%%%%%%%%%%%%%%%%%%%%%%%%%%%%%%%%%%%%%%%%%

\section{System Model and Removing ISI}
This section consists of two subsections. In the first subsection, the system model for an optical communication system using Hermitian symmetry to generate the real valued OFDM waveform for data transmission over a multipath channel is given in matrix form.
The proposed schemes to construct OFDM waveform is introduced in the second subsection using the same system model. 
It is also shown that CRIP and Hermitian symmetry methods can remove ISI using a single tap equalizer with similar way.

\subsection{Hermitian Symmetry Method}
\label{s:hermitian_symmetry}
Assume that $\left[{(N/2)-1} \right]\text{log}_{2}\text{M}^2$ bits data will be transmitted, where $N$ is the number of subcarriers and M represents modulation depth of a complex valued modulation at one dimension. After $\text{M}^2\text{--ary}$ complex mapping, data symbols can be given in vector form as $\underline{x_H}=\left[x_0 \ x_1 \ \ldots \ x_{(N/2)-2} \right]^T$. In this method, before IFFT process, complex valued data symbols are arranged according to structure of Hermitian symmetry as given follows, 
\begin{equation} \label{d:hermitian}
\underline{S_H}=\left[0 \ x_0 \ x_1 \ \ldots \ x_{(N/2)-2} \ 0 \ x^{*}_{(N/2)-2} \ \ldots \ x^{*}_1 \ x^{*}_0 \right]^T
\end{equation}
where $(.)^*$ denotes complex conjugate operator. $\underline{S_H}$ is a column vector in the form of $\underline{S_H}=\left[s_0 \ s_1 \ \ldots \ s_{N-1} \right]^T$, $s_{0}=s_{N/2}=0$ and $s_{k}=s^{*}_{N-k}=x_{k-1}$ for $k=1,2,\ldots,(N/2)-1$.
The Hermitian symmetrical structure of the $\underline{S_H}$ vector ensures that the IFFT of $\underline{S_H}$ will have only real valued elements.
Note that only $(N/2)-1$ complex valued data symbols can be transmitted via $N$ subcarriers in this way. $\underline{S_{F_{H}}}=\left[S_0 \ S_1 \ \ldots \ S_{N-1} \right]^T$ represents IFFT of $\underline{S_H}$, and it can be calculated as follows, 
\begin{equation} \label{d:SFH}
	\underline{S_{F_{H}}}=\underline{\underline{F^H}} \ \underline{S_H}, 
\end{equation}
where $\underline{\underline{F^H}}$ stands for inverse Fourier transform matrix. $\underline{\underline{F^H}}$ matrix is obtained by $\underline{\underline{F^H}}=\left( \underline{\underline{F^T}} \right)^* $, where $\underline{\underline{F}}$ is Fourier transform matrix and given with Eq.~(\ref{d:fourier}). In the matrix, $W_{N}$ is equal~to~$e^{-\frac{j2\pi}{N}}$.
\begin{align} \label{d:fourier}
&\arraycolsep=3pt
\underline{\underline{F}}=\dfrac{1}{\sqrt{N}}\left[
\begin{array}{@{\hskip1pt}ccccc@{\hskip1pt}}
1		& 1			& 1				& \ldots 	& 1 \\
1		& W_N		& W_N^2			& \ldots 	& W_N^{N-1} \\ 
\vdots	& \vdots	& \vdots 		& \ddots 	& \vdots \\	
1		& W_N^{N-1} & W_N^{2(N-1)} 	& \ldots  	& W_N^{{(N-1)}^2} \\ 	
\end{array}
\right]
\end{align} 
$\underline{S_{F_{H}}}$ is the real valued data symbols vector to be transmitted over a multipath optical channel where ISI is present. 
Assume that number of interfered symbols by ISI is $\mu+1$. 
To prevent interference caused by previous OFDM blocks, $\left[S_{N-1} \ \ldots \ S_{N-\mu} \right]$ vector is added to $\underline{S_{F_{H}}}$ as CP with the length of $\mu$ \cite{Goldsmith}.
After removing CP at the receiver side, received data vector $\underline{Y_H}$ can be given as below,
\begin{equation} \label{d:receivedY2}
\arraycolsep=0.1pt
\underbrace{\left[
\begin{array}{c}
y_{N-1}\\
y_{N-2}\\
\vdots \\
y_{0} \\
\end{array}
\right]}_{\underline{Y_H}}
= 
\underbrace{\left[
	\begin{array}{@{\hskip0pt}ccccccc@{\hskip0pt}}
	h_{0} 	&\ h_{1} 	&\ \ldots &\ h_{\mu} 	&\ \ldots &\ 0 \\
	0 		&\ h_{0} 	&\ \ldots &\ h_{\mu -1} 	&\ \ldots &\ 0 \\
	\vdots 	&\ \vdots 	&\ \ddots &\ \vdots 		&\ \ddots &\ \vdots \\
	h_{2} 	&\ h_{3} 	&\ \ldots &\ 0 	&\ \ldots 	&\ h_{1}\\
	h_{1} 	&\ h_{2} 	&\ \ldots &\ 0      &\ \ldots   &\ h_{0} \\
	\end{array}
	\right]}_{\underline{\underline{H}}}
\underbrace{\left[
\begin{array}{@{\hskip0pt}c@{\hskip0pt}}
S_{N-1}\\
S_{N-2} \\
\vdots \\
S_{0} \\
\end{array}
\right]}_{\underline{S_{F_{H}}}}
+
\underbrace{\left[
\begin{array}{c}
n_{N-1}\\
n_{N-2}\\
\vdots \\
n_{0} \\
\end{array}
\right]}_{\underline{\mathcal{N}}}
\end{equation}
Here, $\underline{\mathcal{N}}$ is additive white Gaussian noise (AWGN) vector and $\underline{\underline{H}}$ denotes channel matrix which is $N \times N$ circulant.
Eq.~(\ref{d:receivedY2}) can be rewritten as,
\begin{align}
	\underline{Y_H}=\underline{\underline{H}} \ \underline{S_{F_{H}}}+\underline{\mathcal{N}}.
\end{align}
Since the signal is going to be transmitted by an LED or laser, a DC bias is added to obtain non-negative valued signal before transmitting.
At the receiver side, the DC bias is filtered and to obtain the transmitted data symbols vector $\underline{S_H}$, FFT process is firstly applied to $\underline{Y_{H}}$ and then, the single tap equalization is performed at OFDM receiver. The output of FFT process can be given as follows,
\begin{align} \label{d:R_H}
	\underline{R_H}=\underline{\underline{F}} \ \underline{Y_H}=\underline{\underline{F}} \ \underline{\underline{H}} \ \underline{S_{F_{H}}}+ \underline{\underline{F}} \ \underline{\mathcal{N}}
\end{align}
A circulant matrix can be written according to the eigenvalue decomposition as $\underline{\underline{H}}=\underline{\underline{F^H}} \ \underline{\underline{\Lambda}} \ \underline{\underline{F}}$. Here, $\underline{\underline{\Lambda}}$ is a diagonal matrix which contains eigenvalues of~$\underline{\underline{H}}$. Using eigenvalue decomposition and (\ref{d:SFH}), Eq. (\ref{d:R_H}) can be arranged as,
\begin{subequations}
\begin{align} \label{d:FFTY}
\underline{R_H}&=\underline{\underline{F}} \ \underline{\underline{F^H}} \ \underline{\underline{\Lambda}} \  \underline{\underline{F}} \ \underline{\underline{F^H}} \ \underline{S_H}+\underline{\underline{F}} \ \underline{\mathcal{N}}& \\
&=\underline{\underline{\Lambda}} \ \underline{S_H} + \underline{\underline{F}} \ \underline{\mathcal{N}}&
\end{align}
\end{subequations}
After the FFT process, $\underline{R_H}$ vector is multiplied by a single tap equalizer matrix $\underline{\underline{\Lambda^{-1}}}$ and thus the equalizer output $\underline{RR_H}$ is obtained as follows,
\begin{subequations}
\begin{align} \label{d:EqualizationR}
\underline{RR_H}&=\underline{\underline{\Lambda^{-1}}} \ \underline{\underline{\Lambda}} \ \underline{S_H} + \underline{\underline{\Lambda^{-1}}} \ \underline{\underline{F}} \ \underline{\mathcal{N}}  \\
&=\underline{S_H} + \underline{\underline{\Lambda^{-1}}} \ \underline{\underline{F}} \ \underline{\mathcal{N}}&
\end{align}
\end{subequations}
where $\underline{S_H}$ is complex valued data symbols in Hermitian symmetric form and $\underline{\underline{\Lambda^{-1}}} \ \underline{\underline{F}} \ \underline{\mathcal{N}}$ is complex valued noise component.
As it can be seen, $\underline{S_H}$ doesn't have any ISI component. 

%%%%%%%%%%%%%%%%%%%%%%%%%%%%%%%%%%%%%%%%%%%%%%%%%%%%%%%%%%%%%%%%%%%%%%%%%%%%%%%%%%%%%%%%%%%%

\subsection{Proposed CRIP Method}
\label{s:proposed_method}
\begin{figure*}[ht]
	\centering
	\includegraphics[scale=0.37]{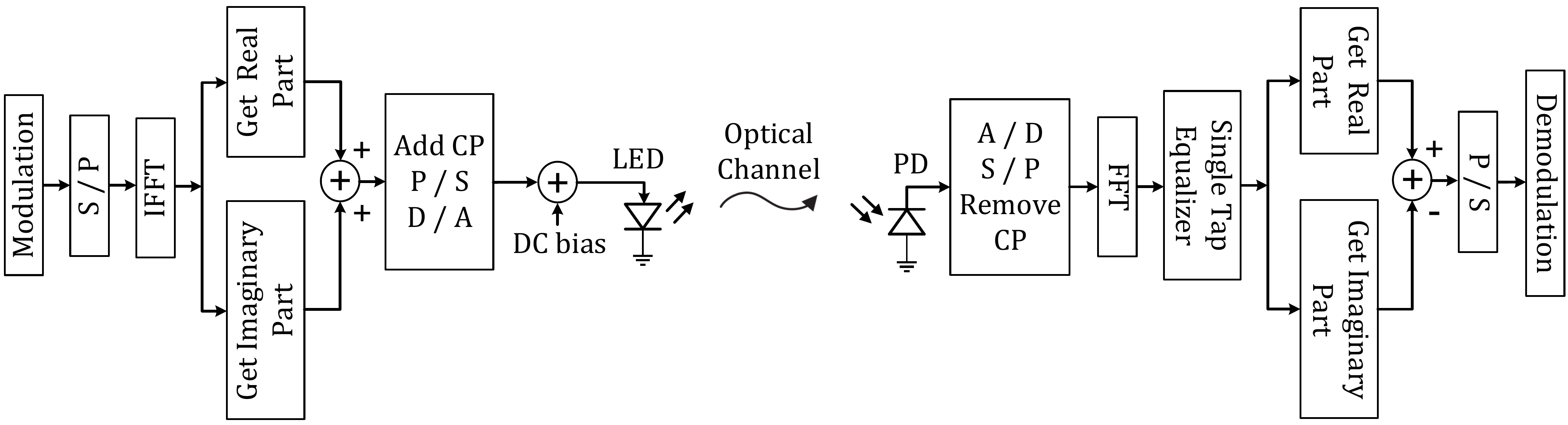}
	\caption{Receiver and transmitter block diagram of a VLC system using the proposed CRIP method. Real and imaginary parts combined at the output of IFFT (E-CRIP).}
	\label{f:proposedblock}
\end{figure*}

%sum of real and imaginary parts (CRIP)
Block diagram of E-CRIP method is given in Fig.~\ref{f:proposedblock}.
The method works on real valued symbols such as M-PAM or M-ASK symbols.
At the transmitter side, IFFT of $N$ real valued symbols is calculated. Then, real and imaginary parts of obtained complex  valued symbols at the IFFT output are summed to get $N$ real valued symbols. After adding CP and DC bias, data vector is transmitted.
At the receiver side, following the DC bias and CP removing, real and imaginary parts of FFT outputs are subtracted from  each other to recover $N$ original real valued data symbols.

\begin{figure}[h]
	\centering
	\includegraphics[scale=0.059]{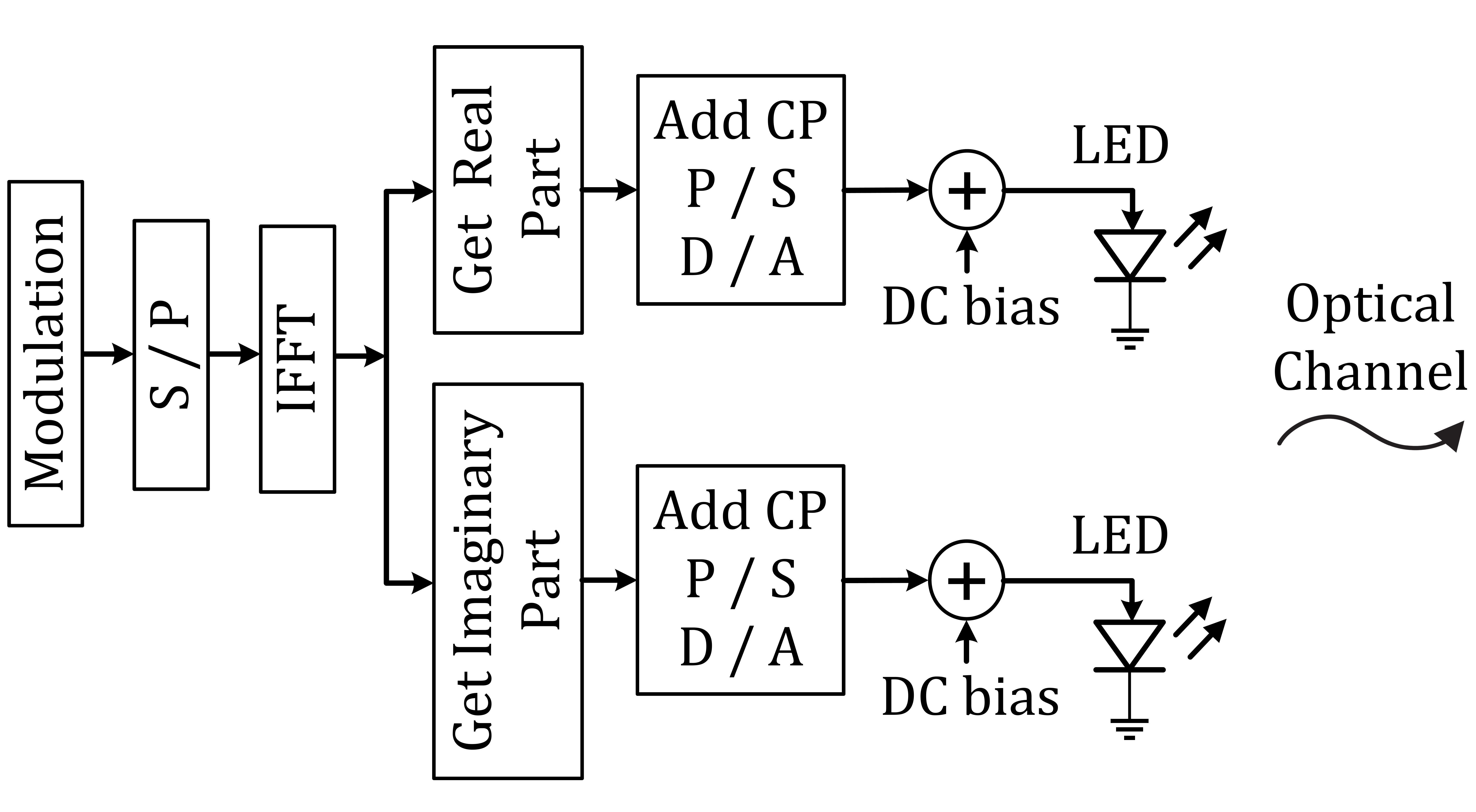}
	\caption{Transmitter block diagram of a VLC system using the proposed CRIP method. Real and imaginary parts of IFFT are combined optically (O-CRIP).}
	\label{f:proposedblock_air}
\end{figure}
%$\underline{S_P}=\left[s_0 \ s_1 \ \ldots \ s_{N-1} \right]^T$ are column \text{matrix} form of subcarriers carried the data symbols

Assume $\underline{x_P}=\left[x_0 \ x_1 \ \ldots \ x_{N-1} \right]^T$ vector has real valued data symbols produced by real M--ary mapping process. Since all OFDM subcarriers are used to carry the elements of $\underline{x_P}$ in CRIP, the subcarriers are arranged as $\underline{S_P}=\left[x_0 \ x_1 \ \ldots \ x_{N-1} \right]^T$. Since IFFT of $\underline{S_P}$ will be complex valued, the obtained signal $\underline{S_{F}}$ after IFFT process can be written as given follows,
\begin{align} \label{d:IFFTproposed}
\underline{S_F}=\underline{\underline{F^H}} \ \underline{S_P}=\underline{S_{FR}}+j\underline{S_{FI}}
\end{align}
In this equation, $\underline{S_{FR}}$ and $\underline{S_{FI}}$ denote real and imaginary parts of $\underline{S_{F}}$, respectively. To obtain real OFDM signal in E-CRIP method, $\underline{S_{FR}}$ and $\underline{S_{FI}}$ are summed, i.e $\underline{S_{F_{P}}}= \underline{S_{FR}}+\underline{S_{FI}}$, and then the real valued $\underline{S_{F_{P}}}$ signal is transmitted over an optical channel where ISI is present. 
CP is also added to prevent interference caused by previous OFDM blocks.
At the receiver side, after removing CP, received vector $\underline{Y_{E-CRIP}}$ for E-CRIP can be given as follows,
\begin{subequations}
\begin{align} \label{d:E-CRIP-a}
\underline{Y_{E-CRIP}}&=\underline{\underline{H}} \ \underline{S_{F_{P}}}+\underline{\mathcal{N}} \\
\label{d:E-CRIP-b}
&=\underline{\underline{H}}\left( \underline{S_{FR}}+\underline{S_{FI}} \right)+\underline{\mathcal{N}}.
\end{align}
\end{subequations}

On the other hand, optical wireless communication systems that can have more than one light source, such as LED-based systems, the combination of real and imaginary parts can also be made optically. As seen in Fig. \ref{f:proposedblock_air}, in the O-CRIP method, the real and imaginary parts feed separate LEDs and the combining is done optically. In this case, received vector $\underline{Y_{O-CRIP}}$ for O-CRIP method can be given as follows,
\begin{subequations}
\begin{align} \label{d:O-CRIP-a}
	\underline{Y_{O-CRIP}}&=\underline{\underline{H}} \ \underline{S_{FR}}+ \underline{\underline{H}} \ \underline{S_{FI}} +\underline{\mathcal{N}} \\
	\label{d:O-CRIP-b}
	&=\underline{\underline{H}}\left( \underline{S_{FR}}+\underline{S_{FI}} \right)+\underline{\mathcal{N}}.
\end{align}
\end{subequations}
Since (\ref{d:E-CRIP-b}) and (\ref{d:O-CRIP-b}) are equal, received vector can be denoted by $\underline{Y_P}$ for both CRIP methods and it can be given as follows,
\begin{align} 
	\underline{Y_P}=\underline{\underline{H}} \ \underline{S_{F_{P}}}+\underline{\mathcal{N}}.
\end{align}
Output of FFT process can be given and rearranged as follows,
\begin{subequations}
\begin{align}\label{d:FFTproposedY}
\underline{R_P}&= \underline{\underline{F}} \ \underline{\underline{H}} \ \underline{S_{F_{P}}}+\underline{\underline{F}} \ \underline{\mathcal{N}}&  \\
&=\underline{\underline{F}} \ \underline{\underline{F^H}} \ \underline{\underline{\Lambda}} \ \underline{\underline{F}} \  \underline{S_{F_{P}}} +\underline{\underline{F}} \ \underline{\mathcal{N}}&  \\
&=\underline{\underline{\Lambda}} \ \underline{\underline{F}} \ \underline{S_{F_{P}}}+\underline{\underline{F}} \ \underline{\mathcal{N}}.&
\end{align}
\end{subequations}
After FFT process, the single tap equalization process is performed as given in Eq.~(\ref{d:EqualizationProposedR}).
\begin{subequations}
\begin{align} \label{d:EqualizationProposedR}
\underline{RR}&=\underline{\underline{\Lambda^{-1}}} \ \underline{\underline{\Lambda}} \ \underline{\underline{F}} \ \underline{S_{F_{P}}} +\underline{\underline{\Lambda^{-1}}} \ \underline{\underline{F}} \ \underline{\mathcal{N}}  \\
&=\underline{\underline{F}} \ \left( \underline{S_{FR}}+\underline{S_{FI}} \right)+\underline{\underline{\Lambda^{-1}}} \ \underline{\underline{F}} \ \underline{\mathcal{N}}&
\end{align}
\end{subequations}
According to CRIP schema, to be able to obtain the transmitted data symbols vector $\underline{S_P}$, the imaginary part of $\underline{RR}$ is subtracted from its real part given as follows,
%$\left( \underline{S_{FR}}+\underline{S_{FI}} \right)$ signal is transmitted. Therefore, to be able to obtain the transmitted data symbols vector $\underline{S_P}$ in the receiver side, the imaginary part of $\underline{RR}$ subtracts from its real part as given with Equ.~(\ref{d:ProposedSub}). 
\begin{subequations}
\begin{align}%\label{d:ProposedSub}
\underline{RR_P}&= Re\left\lbrace \underline{\underline{F}} \left( \underline{S_{FR}}+\underline{S_{FI}} \right) \right\rbrace   + Re\{\underline{\underline{\Lambda^{-1}}} \ \underline{\underline{F}} \ \underline{\mathcal{N}} \} \nonumber \\
&-Im\left\lbrace \underline{\underline{F}} \left( \underline{S_{FR}}+\underline{S_{FI}} \right) \right\rbrace- Im\{\underline{\underline{\Lambda^{-1}}} \ \underline{\underline{F}} \ \underline{\mathcal{N}} \}&  \\
\label{d:ProposedSub}
&=Re\{\underline{\underline{F}} \ \underline{S_{FR}} \}+Re\{\underline{\underline{F}} \ \underline{S_{FI}} \} +Re\{\underline{\underline{\Lambda^{-1}}} \ \underline{\underline{F}} \ \underline{\mathcal{N}} \} \nonumber \\&-Im\{\underline{\underline{F}} \ \underline{S_{FR}} \} -Im\{\underline{\underline{F}} \ \underline{S_{FI}} \} -Im\{\underline{\underline{\Lambda^{-1}}} \ \underline{\underline{F}} \ \underline{\mathcal{N}} \}
\end{align}
\end{subequations}
In the equation, $Re\{.\}$ and $Im\{.\}$ are real and imaginary parts operator and $\underline{RR_P}$ symbolizes $( Re\{\underline{RR} \}-Im\{\underline{RR} \} )$.
Here, since the $\underline{S_P}$ vector is real valued, $\underline{S_{FR}}$ and $\underline{S_{FI}}$ are even and odd functions, respectively as given in Appendix. According to this, real and imaginary parts of ${ \underline{\underline{F}} \ \underline{S_{FR}} }$ and ${ \underline{\underline{F}} \ \underline{S_{FI}} }$ can be given as follows (see Appendix),
\begin{align}\label{d:evenoddfunc1}
&Re\{\underline{\underline{F}} \ \underline{S_{FI}} \}=0 \ &   \\
&Im\{\underline{\underline{F}} \ \underline{S_{FR}} \}=0 \ &   \\
&Re\{\underline{\underline{F}} \ \underline{S_{FR}} \}=\underline{\underline{F}} \ \underline{S_{FR}} \ &  \\ 
\label{d:evenoddfunc2}
&Im\{\underline{\underline{F}} \ \underline{S_{FI}} \}=-j \ \underline{\underline{F}} \ \underline{S_{FI}}.&
\end{align}
Eq.~(\ref{d:ProposedSub}) is rearranged by considering the expressions in~(\ref{d:evenoddfunc1})~--~(\ref{d:evenoddfunc2}) and Eq.~(\ref{d:ProposedSub2-a}) is obtained as follows,
\begin{subequations}
\begin{align}\label{d:ProposedSub2-a}
\underline{RR_P}&=\underline{\underline{F}} \ \underline{S_{FR}}+ 0 - 0 - \left(-j \ \underline{\underline{F}} \ \underline{S_{FI}} \right) \nonumber \\&
+Re\{\underline{\underline{\Lambda^{-1}}} \ \underline{\underline{F}} \ \underline{\mathcal{N}} \}- Im\{\underline{\underline{\Lambda^{-1}}} \ \underline{\underline{F}} \ \underline{\mathcal{N}} \}&  \\
\label{d:ProposedSub2-b}
&=\underline{\underline{F}} \ \left( \underline{S_{FR}} + j \ \underline{S_{FI}} \right) \nonumber \\& +Re\{\underline{\underline{\Lambda^{-1}}} \ \underline{\underline{F}} \ \underline{\mathcal{N}} \}- Im\{\underline{\underline{\Lambda^{-1}}} \ \underline{\underline{F}} \ \underline{\mathcal{N}} \}
\end{align}
\end{subequations}
Eq.~(\ref{d:ProposedSub2-b}) is rearranged by considering (\ref{d:IFFTproposed}) and the transmitted data symbols vector $\underline{S_P}$ is obtained as, 
\begin{subequations}
\begin{align}
	\label{d:ProposedSub3-a}
	\underline{RR_P}&=\underline{\underline{F}} \ \underline{\underline{F^H}} \ \underline{S_P} +Re\{\underline{\underline{\Lambda^{-1}}} \ \underline{\underline{F}} \ \underline{\mathcal{N}} \}- Im\{\underline{\underline{\Lambda^{-1}}} \ \underline{\underline{F}} \ \underline{\mathcal{N}} \}  \\
	\label{d:ProposedSub3-b}
	&=\underline{S_P} +Re\{\underline{\underline{\Lambda^{-1}}} \ \underline{\underline{F}} \ \underline{\mathcal{N}} \}- Im\{\underline{\underline{\Lambda^{-1}}} \ \underline{\underline{F}} \ \underline{\mathcal{N}} \} 
\end{align}
\end{subequations}
As a result, $\underline{S_P}$ can be obtained without ISI component but with noise. This result proofs that CRIP methods can eliminate ISI as in Hermitian symmetry method.

\section{Analysis of Clipping Noise}

In this section, clipping noise occurred in the presence of signal clipping in the communication system for E-CRIP, O-CRIP and Hermitian symmetry methods is analyzed.
The analyzes show that the clipping noise occurred in E-CRIP and Hermitian symmetry methods is equal to each other, while the clipping noise occurred in the O-CRIP method is significantly less than others.

%-------------------------------------
\subsection{Clipping Noise Power of a Gaussian Vector}
In the literature, OFDM frames with 64 and larger subcarriers are often considered to be Gaussian according to the central limit theorem (CLT) \cite{dimitrov2012clipping, Gauss}.
In this subsection, we assumed that a real valued zero-mean Gaussian signal is applied to an LED after DC biasing, and the biased signal is clipped at independent
bottom and top levels of the LED's dynamic range. Let $\underline{S_{FX}}$ be real valued Gaussian vector with zero-mean and $\sigma_{x}^{2}$ variance, which may represent $\underline{S_{F_H}}$, $\underline{S_{FR}}$, $\underline{S_{FI}}$ or $\left(\underline{S_{FR}}+\underline{S_{FI}}\right)$. Note that since all these vectors are obtained after IFFT based on Hermitian symmetry or proposed methods, they are real valued vector and can be assumed Gaussian with zero-mean but different variances. Let $\underline{S_{FX_C}}$ be clipped version of $\underline{S_{FX}}$ and it can be given as follows,
\begin{align} \label{d:S_FXC}
	\underline{S_{FX_C}}(k)=\begin{cases}
		\underline{S_{FX}}(k), \quad B\leq \underline{S_{FX}}(k)\leq T\\
		B, \quad \quad \quad \quad \underline{S_{FX}}(k) < B \\
		T, \quad \quad \quad \quad \underline{S_{FX}}(k) > T
	\end{cases}
\end{align}

\noindent Here $B$ and $T$ are defined as $B=v_{th}-v_{dc}$ and $T=v_{st}-v_{dc}$, where $v_{th}$, $v_{st}$ and $v_{dc}$ stand for bottom and top levels of a LED's dynamic range, and DC bias voltage level, respectively, and $k$ is the index of vectors. 
The clipping noise component can be expressed as,
\begin{align}
	\underline{\mathcal{N}_{X_C}}=\underline{S_{FX}}-\underline{S_{FX_C}}.
\end{align}
The power of the clipping noise,  $P_{C\mathcal{N},X}$, which is second moment of the clipping noise, can be calculated as follows,
\begin{align}
	P_{C\mathcal{N},X}=E\left[\left( \underline{S_{FX}}(k)-  \underline{S_{FX_C}}(k) \right)^{2}   \right].
\end{align}
Here $E[.]$ is the expected value operator. $P_{C\mathcal{N},X}$ can be derived as follows,
\begin{subequations}
\begin{align} 
	P_{C\mathcal{N},X}&=E\left[\underline{S_{FX}}(k)^{2}+\underline{S_{FX_C}}(k)^{2}-2\underline{S_{FX}}(k)  \underline{S_{FX_C}}(k) \right] \\
	\label{d:P_ClipNoise-b}
	&=E\left[\underline{S_{FX}}(k)^{2}\right]+E\left[\underline{S_{FX_C}}(k)^{2}\right] \nonumber \\
	&-\;2E\left[\underline{S_{FX}}(k) \;\underline{S_{FX_C}}(k)\right].
\end{align}
\end{subequations}
The clipping noise power, $P_{C\mathcal{N},X}$, consists of three expected value terms. To derive it, these three terms must be determined. The first term, $E\left[\underline{S_{FX}}(k)^{2}\right]$, is equal to $\sigma_{x}^{2}$, since $\underline{S_{FX}}$ is defined above as real valued Gaussian vector with zero-mean and $\sigma_{x}^{2}$ variance. Below,  $E\left[\underline{S_{FX_C}}(k)^{2}\right]$ and $E\left[\underline{S_{FX}}(k) \;\underline{S_{FX_C}}(k)\right]$ terms will be derived.
Using Eq. (\ref{d:S_FXC}), $\underline{S_{FX_C}}(k)^{2}$ can be written as follows,
\begin{align}
	\underline{S_{FX_C}}(k)^{2}=\begin{cases}
		\underline{S_{FX}}(k)^{2}, \quad B\leq \underline{S_{FX}}(k)\leq T\\
		B^{2}, \quad \quad \quad \quad \underline{S_{FX}}(k) < B \\
		T^{2}, \quad \quad \quad \quad \underline{S_{FX}}(k) > T
	\end{cases}
\end{align}

\noindent Then, the second moment of $\underline{S_{FX_C}}(k)$ can be written by using conditional expected values as in Eq.~(\ref{d:E_SFXC}). 
\begin{figure*}
\begin{align} \label{d:E_SFXC}
	E\left[\underline{S_{FX_C}}(k)^{2}\right]&=E\left[\underline{S_{FX}}(k)^{2} \mid B\leq \underline{S_{FX}}(k)\leq T \right] Pr\left(B\leq \underline{S_{FX}}(k)\leq T\right) \nonumber \\
	&+B^{2}Pr\left(\underline{S_{FX}}(k) < B\right)+T^{2}Pr\left(\underline{S_{FX}}(k) > T\right)  
\end{align}
\end{figure*}
Here, $Pr(.)$ stands for probability operator and
$E\left[\underline{S_{FX}}(k)^{2} \mid B\leq \underline{S_{FX}}(k)\leq T \right]$ is the conditional expected value of $\underline{S_{FX}}(k)^{2}$ with the condition of $\underline{S_{FX}}(k)$ falling into $\left[B, T\right]$ interval. Note that $( B\leq \underline{S_{FX}}(k)\leq T )$ has the doubly truncated Gaussian distribution because $\underline{S_{FX}}(k)$ is defined as real valued Gaussian vector. The conditional expected value can be computed as in Eq.~(\ref{d:E_SFX2_BT}) since $\underline{S_{FX}}(k)$ has zero mean and variance $\sigma_{x}^{2}$ \cite{Truncated1, Truncated2}.
\begin{figure*}
\begin{align} \label{d:E_SFX2_BT}
	E\left[\underline{S_{FX}}(k)^{2} \mid \left( B\leq \underline{S_{FX}}(k)\leq T\right) \right]=\sigma_{x}^{2}+\sigma_{x}^{2}\frac{\phi^{'}\left(\frac{T}{\sigma_{x}}\right)-\phi^{'}\left(\frac{B}{\sigma_{x}}\right)}{Q\left(\frac{B}{\sigma_{x}}\right)-Q\left(\frac{T}{\sigma_{x}}\right)}
\end{align}
\end{figure*}
In this equation, $Q(x)=\frac{1}{\sqrt{2\pi}} \int_{x}^{\infty}e^{\frac{-t^{2}}{2}}dt $, $\phi(x)=\frac{1}{\sqrt{2\pi}}e^{(\frac{-x^{2}}{2})}$ and $\phi^{'}(x)=-\frac{x}{\sqrt{2\pi}}e^{(\frac{-x^{2}}{2})}=-x\phi\left(x\right)$ which is derivative of $\phi(x)$. With similar way, related probability value can be computed as follows,
\begin{align} \label{d:Pr_SFX_BT}
	Pr\left(B\leq \underline{S_{FX}}(k)\leq T\right)=Q\left(\frac{B}{\sigma_{x}}\right)-Q\left({\frac{T}{\sigma_{x}}}\right)  
\end{align}
On the other hand, $(\underline{S_{FX}}(k) < B)$ and $(\underline{S_{FX}}(k) > T)$ have the lower truncated and the upper truncated Gaussian distribution, respectively. So, their probability can be computed as follows,
\begin{align}\label{d:Pr_SFX_B}
	Pr\left(\underline{S_{FX}}(k) < B\right)=1-Q\left(\frac{B}{\sigma_{x}}\right) 
\end{align}
\begin{align} \label{d:Pr_SFX_T}
	Pr\left(\underline{S_{FX}}(k) > T\right)=Q\left(\frac{T}{\sigma_{x}}\right) 
\end{align}
Then, the second term of Eq. (\ref{d:P_ClipNoise-b}) can be written by combining (\ref{d:E_SFX2_BT}), (\ref{d:Pr_SFX_BT}), (\ref{d:Pr_SFX_B}) and (\ref{d:Pr_SFX_T}). 
To derive the third expected value term of Eq. (\ref{d:P_ClipNoise-b}), $\underline{S_{FX}}(k) \; \underline{S_{FX_C}}(k)$ can be written as follows,
\begin{align}
	\underline{S_{FX}}(k) \; \underline{S_{FX_C}}(k)=\begin{cases}
		\underline{S_{FX}}(k)^{2}, \quad B\leq \underline{S_{FX}}(k)\leq T\\
		B\underline{S_{FX}}(k),  \; \quad \underline{S_{FX}}(k) < B \\
		T\underline{S_{FX}}(k),  \; \quad \underline{S_{FX}}(k) > T
	\end{cases} 
\end{align}

\noindent Then, the expected value of $\underline{S_{FX}}(k) \; \underline{S_{FX_C}}(k) $ can be computed as in Eq.~(\ref{d:33}).
\begin{figure*}
\begin{align} \label{d:33}
	E\left[\underline{S_{FX}}(k)\underline{S_{FX_C}}(k)\right]&=E\left[\underline{S_{FX}}(k)^{2} \mid B\leq \underline{S_{FX}}(k)\leq T \right] Pr\left(B\leq \underline{S_{FX}}(k)\leq T\right) \nonumber \\
	&+B\; E\left[\underline{S_{FX}}(k)\mid \underline{S_{FX}}(k) < B\right]Pr\left(\underline{S_{FX}}(k) < B\right) \nonumber \\
	&+T\; E\left[\underline{S_{FX}}(k)\mid \underline{S_{FX}}(k) > T\right]Pr\left(\underline{S_{FX}}(k) > T\right)  
\end{align}
\end{figure*}
%\noindent Assuming the $\underline{S_{FX}} $ as ergodic $E\left[\underline{S_{FX}}(k)^{2} \right] $ should be equal to $\sigma_{x}^{2} $ .
Here, all multiplication terms have already derived except $E[\underline{S_{FX}}(k)\mid \underline{S_{FX}}(k) < B] $ and $E[\underline{S_{FX}}(k)\mid \underline{S_{FX}}(k) > T] $. Since these values  have the lower truncated and the upper truncated Gaussian distribution, respectively, they can be calculated as follows \cite{Truncated1, Truncated2},
\begin{align}
	E\left[\underline{S_{FX}}(k)\mid \underline{S_{FX}}(k) < B\right]=-\sigma_{x}\frac{\phi\left(\frac{B}{\sigma_{x}}\right)}{1-Q\left(\frac{B}{\sigma_{x}}\right)}  
\end{align}
\begin{align}
	E\left[\underline{S_{FX}}(k)\mid \underline{S_{FX}}(k) > T\right]=\sigma_{x}\frac{\phi\left(\frac{T}{\sigma_{x}}\right)}{Q\left(\frac{T}{\sigma_{x}}\right)} 
\end{align}
Finally variance of clipping noise can be written as in (\ref{d:36}).
\begin{figure*}
\begin{align} \label{d:36}
	P_{C\mathcal{N},X}&=\sigma_{x}^{2}-\sigma_{x}^{2}\left(Q\left(\frac{B}{\sigma_{x}}\right)-Q\left({\frac{T}{\sigma_{x}}}\right)+\frac{B}{\sigma_{x}}\phi\left(\frac{B}{\sigma_{x}}\right)-\frac{T}{\sigma_{x}}\phi\left(\frac{T}{\sigma_{x}}\right)\right) \nonumber \\
	&+B^{2}\left(1-Q\left(\frac{B}{\sigma_{x}}\right)\right)+T^{2}Q\left(\frac{T}{\sigma_{x}}\right) +2B\sigma_{x}\phi\left(\frac{B}{\sigma_{x}}\right)-2T\sigma_{x}\phi\left(\frac{T}{\sigma_{x}}\right) 
\end{align}
\hrulefill
% The spacer can be tweaked to stop underfull vboxes.
\vspace*{4pt}
\end{figure*}

\subsection{Clipping Noise Power of E-CRIP and Hermitian Symmetry Method}
As stated above, the output of IFFT, $\underline{S_{F_H}}$, in the Hermitian symmetry method can be modeled as real valued Gaussian vector with zero-mean and $\sigma^{2}$ variance. In this case, the clipping noise power, $P_{C\mathcal{N},H}$, of Hermitian symmetry method can be given as follows,
\begin{align}
	P_{C\mathcal{N},H}=P_{C\mathcal{N},X}\Big|_{\sigma_{x}^{2}=\sigma^2}.
\end{align}
On the other hand, in the proposed CRIP methods, the real and imaginary vectors and the sum of these two vectors at the IFFT output, $\underline{S_{FR}}$, $\underline{S_{FI}}$, $\left(\underline{S_{FR}}+\underline{S_{FI}}\right)$ respectively, are zero mean and Gaussian distributed. 
Variance of $\left(\underline{S_{FR}}+\underline{S_{FI}}\right)$ must be $\sigma^{2}$ to have the same transmitter output power that the Hermitian symmetry method has. Then, the clipping noise power of E-CRIP  method $P_{C\mathcal{N},E-CRIP}$ can be given as follows,
\begin{align}
	P_{C\mathcal{N},E-CRIP}=P_{C\mathcal{N},X}\Big|_{\sigma_{x}^{2}=\sigma^2}.
\end{align}

\subsection{Clipping Noise Power of O-CRIP Method}
\noindent 
Since the real and imaginary parts feed separate LEDs in the O-CRIP method, $\underline{S_{FR}}$, $\underline{S_{FI}}$ are clipped separately. Clipping noise for real and imaginary parts can be calculated as below, respectively,
\begin{align}
\underline{\mathcal{N}_{FR_C}}(k) = \left(\underline{S_{FR}}(k)-\underline{S_{FR_C}}(k)\right)
\end{align}
\begin{align}
	\underline{\mathcal{N}_{FI_C}}(k) = \left(\underline{S_{FI}}(k)-\underline{S_{FI_C}}(k)\right)
\end{align}
where $\underline{S_{FR_{C}}}$ and $\underline{S_{FI_{C}}}$ denote clipped version of $\underline{S_{FR}}$ and $\underline{S_{FI}}$, respectively. In this case, the clipping noise power of O-CRIP  method $P_{C\mathcal{N},O-CRIP}$ can be given as follows,
\begin{subequations}
\begin{align}
	P_{C\mathcal{N},O-CRIP}&=E\left[\left(\underline{\mathcal{N}_{FR_C}}(k)+\underline{\mathcal{N}_{FI_C}}(k)\right)^{2}\right]  \\
	&=E\left[\underline{\mathcal{N}_{FR_C}}(k)^{2}\right]+E\left[\underline{\mathcal{N}_{FI_C}}(k)^{2}\right] \nonumber \\
	&+2 E\left[\underline{\mathcal{N}_{FR_C}}(k)\: \underline{\mathcal{N}_{FI_C}}(k)\right]
\end{align}
\end{subequations}
If $\underline{\mathcal{N}_{FR_C}}(k)$ and $\underline{\mathcal{N}_{FI_C}}(k)$ are assumed as independent and have equal moments, $P_{C\mathcal{N},O-CRIP}$ can be calculated as follows,
\begin{align} \label{d:P_ClipNoise_OC}
	P_{C\mathcal{N},O-CRIP}&=2 P_{C\mathcal{N},X}\Big|_{\sigma_{x}^{2}=\frac{\sigma^2}{2}} \nonumber \\
	&+ 2 E\left[\underline{\mathcal{N}_{FR_C}}(k)\right] E\left[\underline{\mathcal{N}_{FI_C}}(k)\right]
\end{align}
Here, since the variance of $\left(\underline{S_{FR}}+\underline{S_{FI}}\right)$ is $\sigma^{2}$ as aforementioned, variances of $\underline{S_{FR}}$ and $\underline{S_{FI}}$ are both equal to $\sigma^{2}/2$. The second term of Eq. (\ref{d:P_ClipNoise_OC}) can be derived as follows,
\begin{align} \label{d:NxN}
	E\left[\underline{\mathcal{N}_{FR_C}}(k)\right] E\left[\underline{\mathcal{N}_{FI_C}}(k)\right] = \left(E\left[\underline{S_{FX_C}}(k)\right]\right)^{2}
\end{align}
Let $\underline{\mathcal{N}_{X_C}}$ represents clipping noise of real or imaginary parts. It is defined as,
\begin{align}
	\underline{\mathcal{N}_{X_C}}=\underline{S_{FX}}-\underline{S_{FX_C}}.
\end{align}
Expected value of $\underline{\mathcal{N}_{X_C}}$ can be given as,
\begin{subequations}
\begin{align}
	E\left[\underline{\mathcal{N}_{X_C}}(k)\right]&=E\left[\underline{S_{FX}}(k)-\underline{S_{FX_C}}(k)\right] \\
	&=E\left[\underline{S_{FX}}(k)\right]-E\left[\underline{S_{FX_C}}(k)\right]
\end{align} 
\end{subequations}
Since $\underline{S_{FX}}$ is zero-mean Gaussian vector {\Large(}$E\left[\underline{S_{FX}}(k)\right]=0${\Large)} and
\begin{align}
	E\left[\underline{\mathcal{N}_{X_C}}(k)\right]=-E\left[\underline{S_{FX_C}}(k)\right]
\end{align}
which proofs the Eq. (\ref{d:NxN}). In this case clipping noise power of O-CRIP method can be given as:
\begin{align}
	P_{C\mathcal{N},O-CRIP}=2 P_{C\mathcal{N},X}\Big|_{\sigma_{x}^{2}=\frac{\sigma^2}{2}}+2 \left(E\left[\underline{S_{FX_C}}(k)\right]\right)^{2}.
\end{align}
In this equation 
$E\left[\underline{S_{FX_C}}(k)\right]$ can be calculated as in (\ref{d:E_SFX_C_k}).
\begin{figure*}
\begin{align}\label{d:E_SFX_C_k}
	E\left[\underline{S_{FX_C}}(k)\right]&=E\left[\underline{S_{FX}}(k) \mid \left( B\leq \underline{S_{FX}}(k)\leq T\right) \right] Pr\left(B\leq \underline{S_{FX}}(k)\leq T\right) \nonumber \\
	&+B\times Pr\left(\underline{S_{FX}}(k) < B\right)+T\times Pr\left(\underline{S_{FX}}(k) > T\right) 
\end{align}
\end{figure*}
In this equation, all terms except $E\left[\underline{S_{FX}}(k) \mid B \leq \underline{S_{FX}}(k) \leq T\right]$ have already been determined above. As mentioned before, $( B\leq \underline{S_{FX}}(k)\leq T )$ has the doubly truncated Gaussian distribution because $\underline{S_{FX}}(k)$ is defined as real valued Gaussian vector. The conditional expected value can be computed as in Eq.~(\ref{d:E_SFX}) since $\underline{S_{FX}}(k)$ has zero mean and variance $\sigma_{x}^{2}$ \cite{Truncated1, Truncated2}.
\begin{figure*}
\begin{align} \label{d:E_SFX}
	E\left[\underline{S_{FX}}(k) \mid B \leq \underline{S_{FX}}(k) \leq T\right] =\left(-\sigma_{x} \frac{\phi\left(\frac{T}{\sigma_{x}}\right)-\phi\left(\frac{B}{\sigma_{x}}\right)}{Q\left(\frac{B}{\sigma_{x}}\right)-Q\left(\frac{T}{\sigma_{x}}\right)}\right) 
\end{align}
\end{figure*}
Then, $E\left[\underline{S_{FX_C}}(k)\right]$ can be written as in Eq.~(\ref{d:E_SFX_C}) by combining (\ref{d:Pr_SFX_BT}), (\ref{d:Pr_SFX_B}), (\ref{d:Pr_SFX_T}) and (\ref{d:E_SFX}).
\begin{figure*}
\begin{align}\label{d:E_SFX_C}
	E\left[\underline{S_{FX_C}}(k)\right]=-\sigma_{x}\left(\phi\left(\frac{T}{\sigma_{x}}\right)-\phi\left(\frac{B}{\sigma_{x}}\right)\right)+B\left(1-Q\left(\frac{B}{\sigma_{x}}\right)\right)+T Q\left(\frac{T}{\sigma_{x}}\right) 
\end{align}
\hrulefill
% The spacer can be tweaked to stop underfull vboxes.
\vspace*{4pt}
\end{figure*}
Finally clipping noise for the proposed system with two LEDs, O-CRIP, can be given as follows,
\begin{align}
	P_{C\mathcal{N},O-CRIP}&=2 P_{C\mathcal{N},X}\Big|_{\sigma_{x}^{2}=\frac{\sigma^2}{2}} \nonumber \\
	&+2 \left(E\left[\underline{S_{FX_C}}(k)\right]_{\sigma_{x}^{2}=\frac{\sigma^2}{2}}\right)^{2}.
\end{align}
%since the variance of $\underline{S_{FR}}$ and $\underline{S_{FI}}$ must be equal to $\sigma^{2}/2$.

\subsection{Analytic and Simulation Results}
\begin{figure}[ht]
	\centering
	\includegraphics[scale=0.5]{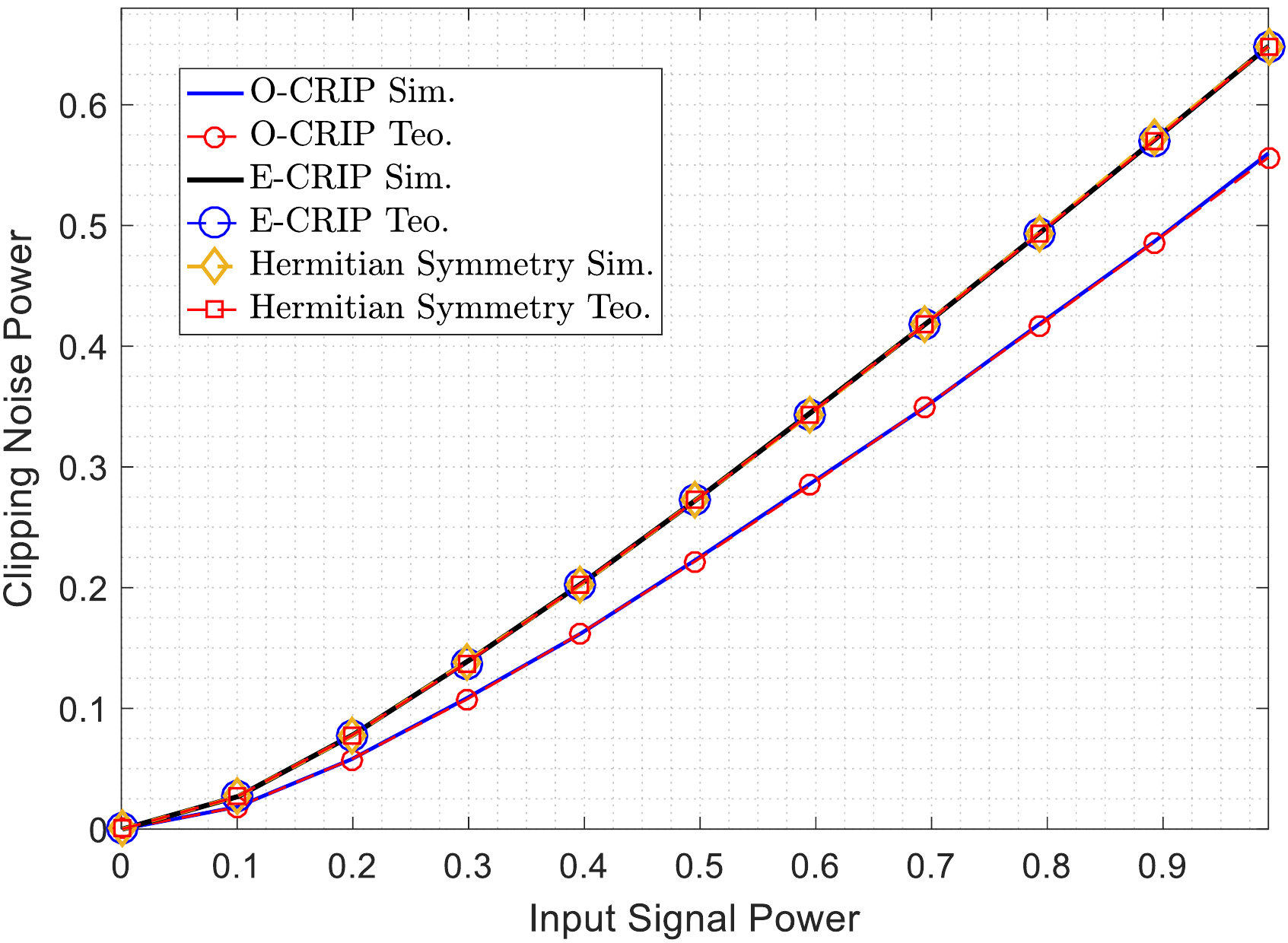}
	\caption{Variation curves of clipping noise power versus input signal power for Hermitian symmetry and proposed CRIP methods.}
	\label{f:ClippingNoise}
\end{figure}

Fig. \ref{f:ClippingNoise} represents the variation curves of clipping noise power, $P_{C\mathcal{N}}$, with respect to input signal power, $\sigma_{x}^{2}$, for Hermitian symmetry and proposed CRIP methods.
Analytic and simulation results validate each other.
For numerical works, a Cree Xlamp XB-H LED chip is considered as a transmitter. The active region of this LED is between 2.65~V~--~3.15~V. Therefore, threshold, saturation and DC bias values are chosen as $v_{th}=2.65$, $v_{st}=3.15$ and $v_{dc}=(2.65+3.15)/2=2.9$, respectively, and $B=-0.25$, $T=0.25$.
As can be seen in the figure, the clipping noise occurred by the O-CRIP method is lower than other methods. Although the figure compares the clipping noise occurred by the methods, it does not show the effect of clipping noise on the performance of the communication system. The comparison of the effect of clipping noise on system error performance that may occur due to the use of off-the-shelf components is given in Section~\ref{s:Numerical Res}. 

%%%%%%%%%%%%%%%%%%%%%%%%%%%%%%%%%%%%%%%%%%%%%%%%%%%%%%%%%%%%%%%%%%%%%%%%%%%%%%%%%%%%%%%%%%%%%%%%

\section{Complexity Analyzes}
\label{s:complexity_analyzes}

In this section, we analyze the computational complexities of O-CRIP, E-CRIP, Hermitian symmetry, and DCT methods.
While CRIP methods perform IFFT operation on real valued symbols, Hermitian symmetry method performs this operation on symbols with complex values. This creates a significant difference in computational complexities of methods. In \cite{vetterli1984simple}, the multiplication and addition processes required for IFFT (or FFT) and DCT are presented in detail.   
Accordingly, we count the computational complexities of IFFT on real valued symbols for the proposed CRIP methods and on complex valued symbols for Hermitian symmetry method. 
The results, including DCT on real valued symbols, are given in Table~\ref{tab:complexity} for only transmitter side. For ease of notation, we define $C_{1}$ and $C_{2}$ values as follows,
\begin{align}
	C_{1}&=(N/2)\times(\text{log}_{2}(N)-3)+2, \\
	C_{2}&=(N/2)\times(3 \text{log}_{2}(N)-5)+4.
\end{align} 
Unlike O-CRIP method, real and imaginary parts of IFFT output are summed to get real valued symbols in E-CRIP. Therefore, E-CRIP method requires extra $N$ additions compared to O-CRIP as shown in the table. Addition to specified operations in Table~\ref{tab:complexity}, both O-CRIP and E-CRIP methods require $N$ subtractions at the receiver side to recover original data, while Hermitian symmetry method needs $(N-2)/2$ complex conjugate operation to get the Hermitian symmetry structure as given in Section~\ref{s:hermitian_symmetry}.

\begin{table}[h]
	\begin{center}
		\caption{Computational complexities comparison of CRIP, Hermitian symmetry and DCT methods}
		\label{tab:complexity}
		\renewcommand{\arraystretch}{1.3}
		\begin{tabular}{|l|c|c|}
			\hline
			\multicolumn{1}{|c|}{\multirow{2}{*}{\textbf{Methods}}} & \multicolumn{2}{c|}{\textbf{Operations}}    \\ \cline{2-3} 
			\multicolumn{1}{|c|}{}                                  & \textit{Multiplication} & \textit{Addition} \\ \hline
			\text{O-CRIP}                                        & $C_{1}$                      		& $C_{2}$                \\ \hline
			\text{E-CRIP}                                         & $C_{1}$                      		&$ C_{2}+N$              \\ \hline
			\text{Hermitian Symmetry}                    & $2C_{1}$                     	& $2C_{2}+2N-4$          \\ \hline
			\text{DCT}                                             & $C_{1}+1.5N-2$               	& $C_{2}+1.5N-3$         \\ \hline
		\end{tabular}
	\end{center}
\end{table}

It is clear from Table~\ref{tab:complexity} that proposed CRIP methods have lower computational complexity than both Hermitian symmetry and DCT methods. However, the numbers of required operations for all discussed methods are depend on number of subcarriers. 
To better understand the differences between the computational complexities of the methods, the numbers of multiplication and addition operations calculated for various $N$ are given in Table~\ref{tab:complexity2}. 
As can be seen, the amount of calculation required by CRIP methods is quite low compared to other methods. 
And also, as the number of subcarriers increases, the difference gets further.
%As can be seen from the Table, when the number of subcarriers increases, the number of operations required by DCT less increases compared to Hermitian symmetry method. Furthermore, the proposed method has less computational complexity than DCT, and also the difference of number of operations required by DCT and CRIP becomes wider as the number of subcarriers increases.
\begin{table}[h]
	\begin{center}
		\caption{Numbers of operations required by all methods for various numbers of subcarriers}
		\label{tab:complexity2}
		\renewcommand{\arraystretch}{1.3}
		\begin{tabular}{|>{\centering\arraybackslash}m{0.4in}|l|>{\centering\arraybackslash}m{0.06in}|>{\centering\arraybackslash}m{0.1in}|>{\centering\arraybackslash}m{0.1in}|>{\centering\arraybackslash}m{0.1in}|>{\centering\arraybackslash}m{0.17in}|>{\centering\arraybackslash}m{0.17in}|>{\centering\arraybackslash}m{0.22in}|}
			\hline
			\multirow{2}{*}{\textbf{Methods}} & \multicolumn{1}{c|}{\multirow{2}{*}{\textbf{Operations}}} & \multicolumn{7}{c|}{\textbf{Number of subcarriers (N)}} \\ \cline{3-9} 
			& \multicolumn{1}{c|}{} &  \textbf{\textit{8}}    &  \textbf{\textit{16}}    &  \textbf{\textit{32}}    &  \textbf{\textit{64}}    &  \textbf{\textit{128}}    &  \textbf{\textit{256}}    &  \textbf{\textit{512}}    \\ \hline
			\multirow{2}{*}{O-CRIP}		& \textit{Multip.} 	 &  2    &    10   &  34     &   98    &    258    &  642      &   1538     \\ \cline{2-9} 
			& \textit{Addition}    &  20    &   60    &    164   &   420    &    1028    &   2436     &  5636      \\ \hline
			\multirow{2}{*}{E-CRIP}             & \textit{Multip.}      &   2   &   10    &    34   &   98    &    258    &   642     &  1538      \\ \cline{2-9} 
			& \textit{Addition}   &  28    &   76    &    196   &  484     &   1156     &    2692    &   6148     \\ \hline
			\multirow{2}{*}{\begin{tabular}[c]{@{}c@{}}Hermitian\\ Symmetry\end{tabular}} & \textit{Multip.}    &  4    &   20    &    68   &   196    &   516     &   1284     &  3076      \\ \cline{2-9} 
			& \textit{Addition}    &  52   &   148    &  388     &  964     &   2308     &   5380     &   12292     \\ \hline
			\multirow{2}{*}{DCT}                  & \textit{Multip.}      &   12   &   32    &  80     &   192    &     448   &    1024    &   2304    \\ \cline{2-9} 
			& \textit{Addition}     &  29    &  81     &   209    &    513   &    1217    &   2817     &    6401    \\ \hline
		\end{tabular}
	\end{center}
\end{table}

%%%%%%%%%%%%%%%%%%%%%%%%%%%%%%%%%%%%%%%%%%%%%%%%%%%%%%%%%%%%%%%%%%%%%%%%%%%%%%%%%%%%%%%%%%%%%%%%%%%

\section{Numerical Results for VLC System}
\label{s:Numerical Res}
In this section, we comparatively evaluate OFDM schemas utilized Hermitian symmetry and proposed CRIP methods in terms of their spectral efficiencies and bit error rate (BER) performance over a VLC channel where ISI is present as an example to optical channels. And also, the impacts of the clipping noise caused by the imperfect front-ends of the VLC system are investigated in detail.

For VLC simulation studies, a model room with multiple LED armatures is considered for both lighting and data transmission purpose. The model room with $8$m x $6$m x $3$m size has six LED armatures which are located on ceiling at locations of ($2$m x $2$m x $3$m), ($4$m x $2$m x $3$m), ($6$m x $2$m x $3$m),  ($2$m x $4$m x $3$m), ($4$m x $4$m x $3$m) and ($6$m x $4$m x $3$m). And, each armature consists of 9 LED chips which are commercially available (Cree Xlamp XB-H). A ray tracing based model introduced in \cite{Lee} is used to obtain the VLC channel impulse response. Also, same materials (for walls, floor and ceiling) and the noise parameters are considered as in \cite{CenkKadir_seamless}. It is assumed that all LED chips simultaneously transmit same data in the transmitter side and a photodetector whose active area and field of view are $1$cm$^2$ and $85^\circ$, respectively, is used at the receiver side.
For a fair comparison, we use real valued M-PAM and complex valued M$^2$-QAM constellations for OFDM systems with CRIP and Hermitian symmetry methods, respectively. Here, M represents the modulation scale. We also assume that OFDM symbols are composed of 64 subcarriers and that the length of the cyclic prefix (CP) is $1/8$ symbol for both systems.

\begin{figure}%[ht]
	\centering
	\includegraphics[scale=0.5]{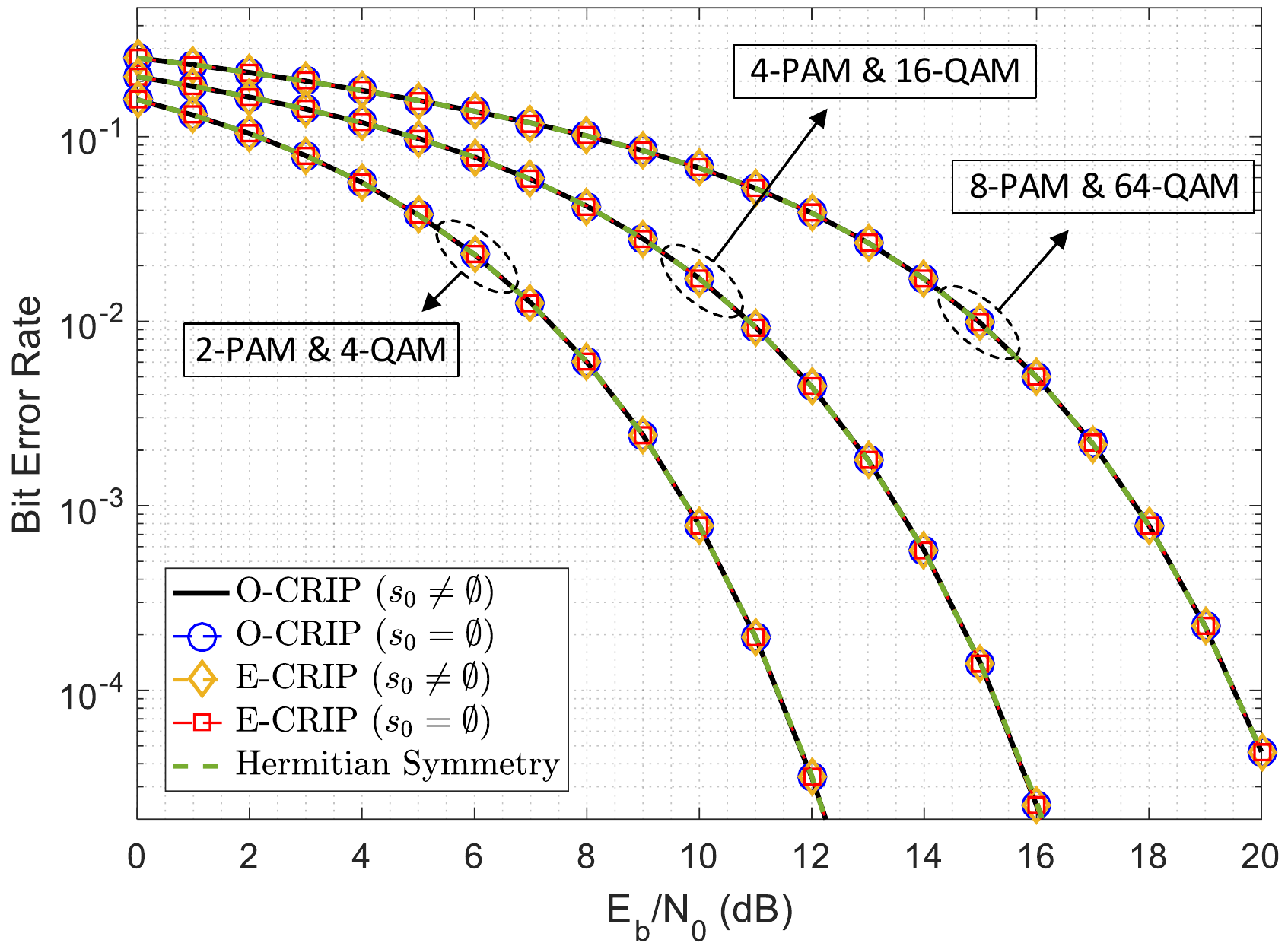}
	\caption{BER performances of OFDM schema utilized Hermitian symmetry and proposed methods for various constellation sizes}
	\label{f:ber}
\end{figure}

Fig.~\ref{f:ber} illustrates the BER performance of O-CRIP, E-CRIP and Hermitian symmetry methods for various M values when signal clipping is ignored. These curves are obtained for 100~MHz bandwidth and $E_{b}/N_{0}$ is calculated in the electrical domain by excluding the DC bias. The simulation results are calculated by averaging BER values obtained by considering that the receiver is located in various point of the model room.
As it can be seen in the figure, BER performance of OFDM system with O-CRIP, E-CRIP and Hermitian symmetry methods looks same.
In the figure, $s_{0}\neq{\emptyset}$ and $s_{0}={\emptyset}$ represent that a symbol is loaded in the first subcarrier of the constructed OFDM frame, and not respectively. 
As mentioned before, to employ the first subcarrier of OFDM frame will lightly differ the mean value of the obtained symbol sequence from zero by $s_0/N$.
However, as seen in the figure, O-CRIP and E-CRIP (with $s_{0}\neq{\emptyset}$ and $s_{0}={\emptyset}$), and Hermitian symmetry methods seem to have the same BER performance. 

Let $R_{H}$, $R_{CRIP_{s_{0}\neq{\emptyset}}}$ and $R_{CRIP_{s_{0}={\emptyset}}}$ represent the peak value of error free transmission rate without considering CP for $N$ subcarriers using optical OFDM with Hermitian symmetry and CRIP methods whose first subcarrier is loaded and is not, respectively. 
These peak values can be calculated as follows,
\begin{equation} 
	R_{H}=\left(\frac{N}{2}-1 \right) \text{log}_{2}\text{M}^{2} = \left(N-2 \right) \text{log}_{2}\text{M}, 
\end{equation}
\begin{equation} \label{d:specefficiency}
	R_{CRIP_{s_{0}\neq{\emptyset}}}=N  \text{log}_{2}\text{M}, 
\end{equation}
\begin{equation} \label{d:specefficiency}
	R_{CRIP_{s_{0}={\emptyset}}}=\left(N-1 \right) \text{log}_{2}\text{M}. 
\end{equation}
It means that proposed CRIP schemas with $s_{0}\neq{\emptyset}$ and $s_{0}={\emptyset}$ can transmit $2 \text{log}_{2}\text{M}$ and $\text{log}_{2}\text{M}$ bits more compared to Hermitian symmetry schema in each OFDM frame, respectively. In case of 100~MHz bandwidth, $N=\text{64}$ and $\text{M}=\text{8}$, the transmission rate is  9.375~Mbps (\%3.2) and 4.6875~Mbps (\%1.6) higher while the peak value of error free transmission rate of Hermitian symmetry is 290.625~Mbps. Note that as the number of $N$ increases, the transmission rate differences will decrease. 

Even if the optimum DC bias level and amplifier gain are determined, the imperfections of the transmitter front-ends may result shifting in DC bias and amplifier gain levels. As a consequence, the undesired impact of clipping noise occurs over the optical communication system. 
As aforementioned, to investigate the impact of the clipping noise over OFDM schema with Hermitian symmetry and proposed CRIP systems, a Cree Xlamp XB-H LED chip is considered as VLC transmitter. 
%The active region of this LED is between 2.65~V~--~3.15~V. 
Determination of the amplifier gain and DC bias level is an important to take full advantage of the active region.
The DC bias level should be middle of the active region of LED.
A lower amplifier gain ensures no clipping, but causes reduction the signal power, i.e. $E_{b}/N_{0}$. Also, higher amplifier gain and wrong DC bias cause the signal clipping and so BER performance degradation. Therefore, to achieve a higher spectral efficiency and better BER performance, optimum amplifier gain and DC bias level need to be used.

\begin{figure}%[ht]
	\centering
	\includegraphics[scale=0.5]{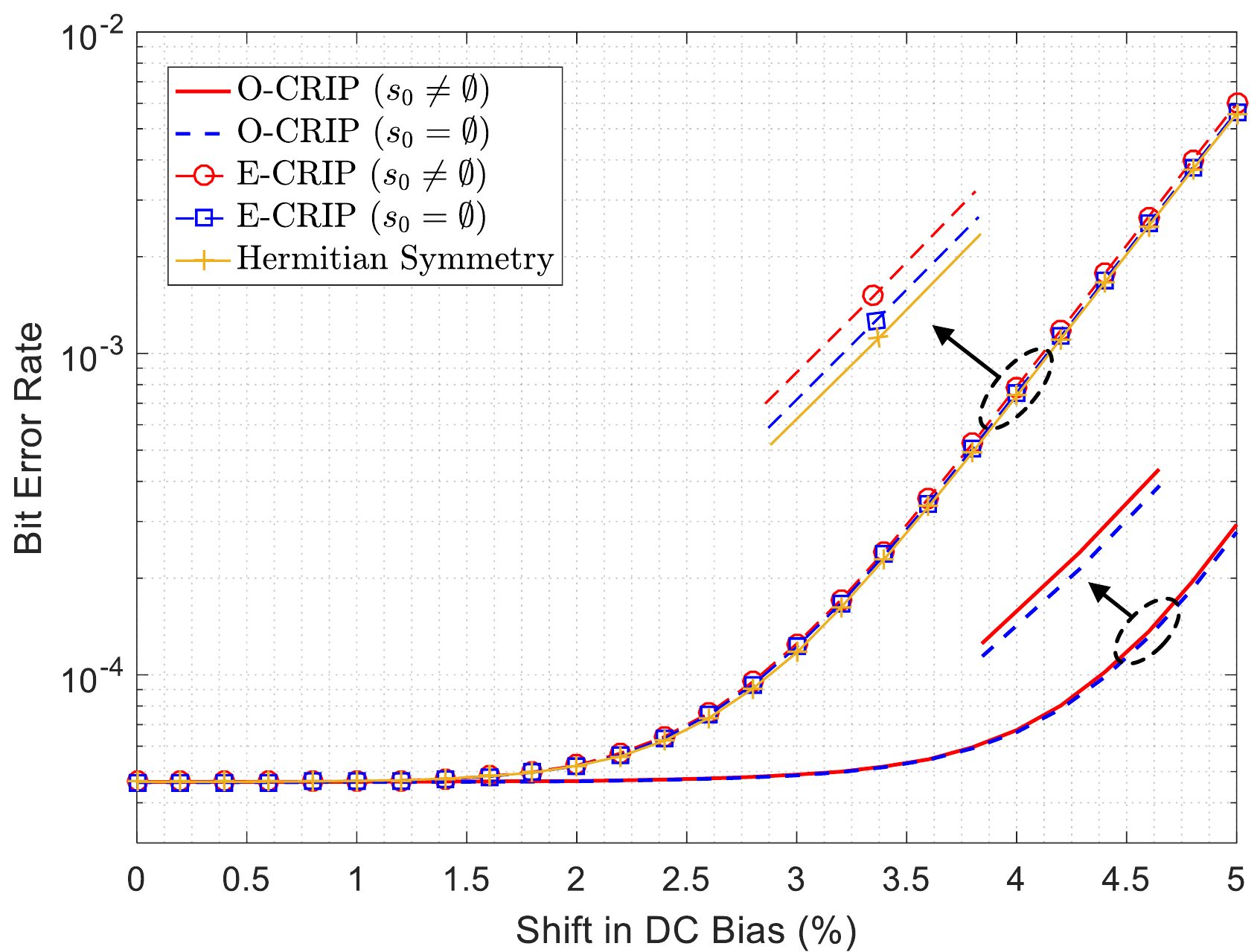}
	\caption{BER performance reduction in Hermitian symmetry and CRIP schemas at $20$ dB due to shift in DC bias}
	\label{f:DC}
\end{figure}

\begin{figure}%[ht]
	\centering
	\includegraphics[scale=0.5]{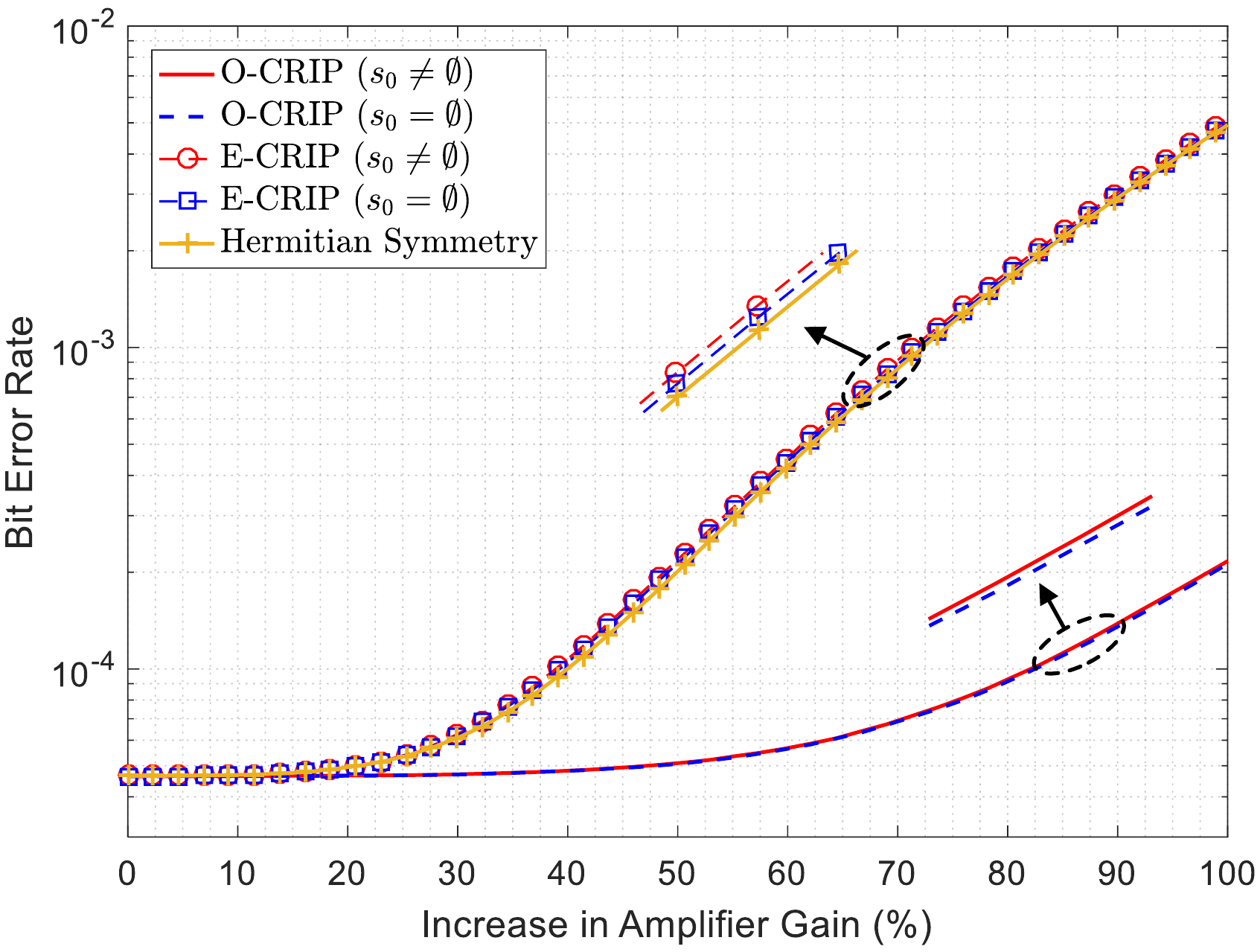}
	\caption{BER performance reduction in Hermitian symmetry and CRIP schemas at $20$ dB due to increase in amplifier gain}
	\label{f:AmpGain}
\end{figure}

Fig.~\ref{f:DC} and Fig.~\ref{f:AmpGain} illustrates the BER performance reduction in Hermitian symmetry and CRIP methods at 20~dB when the DC bias level shifts and the amplifier gain increases, respectively. For a fair comparison, symbol packet energy has been normalized to 1 after modulation process and the optimum amplifier gain has been determined for each schema separately. Also, same DC bias level has been added to the real valued signal at all schemas as $\left(\text{2.65}+\text{3.15}\right)/\text{2}=\text{2.9}$~V As mentioned above, the active region of this LED is between 2.65~V~--~3.15~V. In addition, average noise power at the receiver side has been considered while the impact of~the clipping noise investigates to all schemas. This value is calculated using [17,~Eq. (8) and Table~I] for receiver bandwidth of 100~MHz and the result is averaged over the entire room area. The average noise power $\left(\sigma_{n}^2\right)$ is -98.71~dBm.

From Fig.~\ref{f:DC} and Fig.~\ref{f:AmpGain}, it is clear that the O-CRIP method is much less affected by the clipping noise caused by both the DC bias shift and the increase in the amplifier gain. E-CRIP and Hermitian symmetry methods are almost equally distorted. The figures also show the effect of employing $s_0$ for data transmission on the error performance for the proposed CRIP methods. Using $s_0$ for transmission causes the mean value of the OFDM frame to shift from zero by $s_0/N$. The performance difference caused by the shift of the mean value is shown separately by zooming E-CRIP and O-CRIP curves in the figures. It is seen that the use of the $s_0$ symbol almost does not change the error performance of the system. Note that unlike CRIP methods, the use of the $s_0$ symbol in the Hermitian symmetry method will cause the constructed OFDM frame to have non-real valued symbols.

%%%%%%%%%%%%%%%%%%%%%%%%%%%%%%%%%%%%%%%%%
\section{Conclusion}

In this paper, we propose a novel method for the construction of optical OFDM waveform with IFFT to obtain real valued symbols by combining the real and imaginary parts (CRIP) of IFFT output electrically (E-CRIP) or optically (O-CRIP).
Analytical analysis and simulation works present that the proposed method eliminates ISI with single tap equalizer just like Hermitian symmetry method does. Also, compared to the Hermitian symmetric system, CRIP methods slightly increase the spectral efficiency, significantly reduce the computational complexity and do not effect the error performance. 
In addition, the obtained results show that optical system using O-CRIP method is less affected by clipping noise that may occur due to the imperfections of transmitter front-ends.

%%%%%%%%%%%%%%%%%%%%%%%%%%%%%%%%%%%%%%%%%

\appendix

The application of IDFT to signal $s_k$ is given as follows,
\begin{align}
	S_n=S_{nR}+jS_{nI}=\frac{1}{N}\sum_{k=0}^{N-1}s_k e^{\frac{j2\pi}{N}kn}, \ n=0,1,\ldots,N-1 
\end{align}
where $S_n$ denotes IDFT of $s_k$ signal, $S_{nR}$ and $S_{nI}$ represent real and imaginary parts of $S_n$, respectively. If $s_k$ is real valued, $S_n$ can be written as follows,
\begin{align}
	S_n=\frac{1}{N}\sum_{k=0}^{N-1}s_k cos\left(\frac{2\pi}{N}kn \right)+j\frac{1}{N}\sum_{k=0}^{N-1}s_k sin\left(\frac{2\pi}{N}kn \right). 
\end{align}
$S_{nR}$ and $S_{nI}$ can be given as follows,
\begin{align}
	S_{nR}=\frac{1}{N}\sum_{k=0}^{N-1}s_k cos\left(\frac{2\pi}{N}kn \right), \quad n=0,1,\ldots,N-1 \\
	S_{nI}=\frac{1}{N}\sum_{k=0}^{N-1}s_k sin\left(\frac{2\pi}{N}kn \right), \quad n=0,1,\ldots,N-1 
\end{align}
It is easy to see that since $S_{nR}$ is linear combination of cos function it is even signal and satisfies to the following equality for $n=1,2,\ldots,N-1$.
\begin{subequations}
\begin{align}
	S_{(N-n)R}&=\frac{1}{N}\sum_{k=0}^{N-1}s_k cos\left(\frac{2\pi}{N}k(N-n) \right)  \\
	&=\frac{1}{N}\sum_{k=0}^{N-1}s_k cos\left(\frac{2\pi}{N}nk \right)=S_{nR} 
\end{align}
\end{subequations}
On the other hand $S_{nI}$ is odd signal and satisfies to the following equality for $n=1,2,\ldots,N-1$.
\begin{subequations}
\begin{align}
	S_{(N-n)I}&=\frac{1}{N}\sum_{k=0}^{N-1}s_k sin\left(\frac{2\pi}{N}k(N-n) \right)  \\
	&=-\frac{1}{N}\sum_{k=0}^{N-1}s_k sin\left(\frac{2\pi}{N}nk \right)=-S_{nI} 
\end{align}
\end{subequations}
In this case discrete Fourier transforms (DFT) of $S_{nR}$ and $S_{nI}$ are real and imaginary valued, respectively. They can be given as follows,
\begin{subequations}
\begin{align}
s_{kR}&=\frac{1}{N}\sum_{n=0}^{N-1}S_{nR} \hspace{.12em} cos\left(\frac{2\pi}{N}kn \right)-\frac{j}{N}\sum_{n=0}^{N-1}S_{nR} \hspace{.12em} sin\left(\frac{2\pi}{N}kn \right)  \\
	&=\frac{1}{N}\sum_{n=0}^{N-1}S_{nR} \ cos\left(\frac{2\pi}{N}kn \right), \quad k=0,1,\ldots,N-1 
\end{align}
\end{subequations}
\begin{subequations}
\begin{align}
	s_{kI}&=\frac{1}{N}\sum_{n=0}^{N-1}S_{nI} \ cos\left(\frac{2\pi}{N}kn \right) -\frac{j}{N}\sum_{n=0}^{N-1}S_{nI} \ sin\left(\frac{2\pi}{N}kn \right)  \\
	&=\frac{-J}{N}\sum_{n=0}^{N-1}S_{nI} \ sin\left(\frac{2\pi}{N}kn \right), \quad k=0,1,\ldots,N-1 
\end{align}
\end{subequations}

\ifCLASSOPTIONcaptionsoff
  \newpage
\fi

\end{document}